\documentclass[fleqn,10pt]{wlscirep}

\title{Simulating quantum dynamical phenomena using classical oscillators:\\
Landau-Zener-St\"{u}ckelberg-Majorana interferometry, \\
latching modulation, and motional averaging}

\author[1,2,3]{O.~V.~Ivakhnenko}
\author[1,2,3,*]{S.~N.~Shevchenko}
\author[3,4]{Franco~Nori}

\affil[1]{B. Verkin Institute for Low Temperature Physics and Engineering, Kharkov
61103, Ukraine}
\affil[2]{V.~N. Karazin Kharkov National University, Kharkov 61022, Ukraine}
\affil[3]{Theoretical Quantum Physics Laboratory, RIKEN Cluster for Pioneering Research, Wako-shi, Saitama 351-0198, Japan}
\affil[4]{Physics Department, University of Michigan, Ann Arbor, MI 48109-1040, USA}
\affil[*]{sshevchenko@ilt.kharkov.ua}

\begin{abstract}
A quantum system can be driven by either sinusoidal, rectangular, or noisy
signals. In the literature, these regimes are referred to as Landau-Zener-St%
\"{u}ckelberg-Majorana (LZSM) interferometry, latching modulation, and
motional averaging, respectively. We demonstrate that these pronounced and
interesting effects are also inherent in the dynamics of classical two-state
systems. We discuss how such classical systems are realized using either
mechanical, electrical, or optical resonators. In addition to the fundamental interest of such dynamical phenomena linking classical and quantum physics, we believe that these are attractive for the classical analogue simulation of quantum systems.
\end{abstract}

\begin{document}

\flushbottom
\maketitle

\thispagestyle{empty}

\section{Introduction: Classical-quantum analogies}

\begin{figure}[t]
\centering{\includegraphics[width=0.6 \columnwidth]{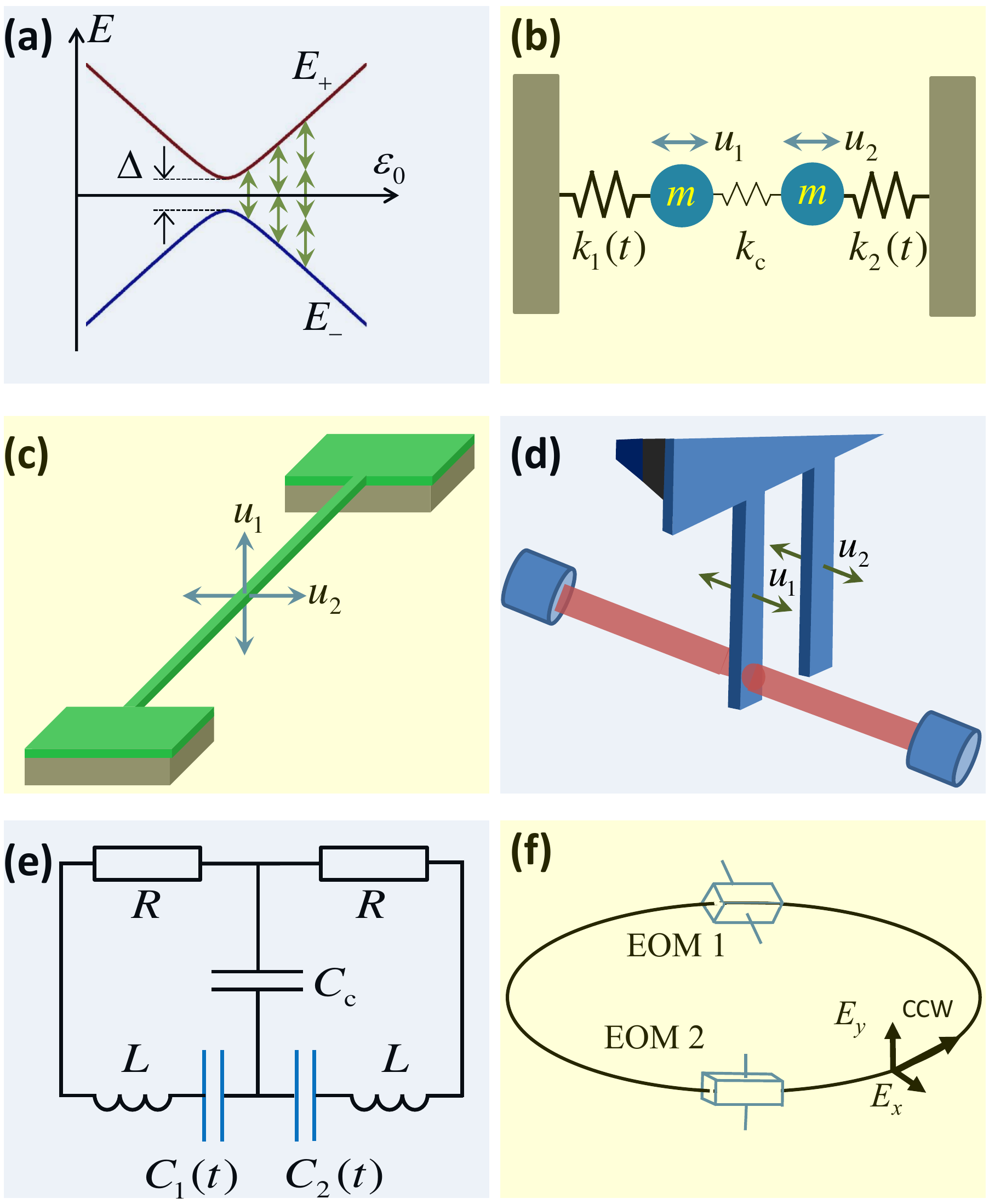}}
\caption{Classical analogues of qubits. In (a) the two qubit eigenenergies $%
E_{\pm }$ are shown to depend on the bias $\protect\varepsilon _{0}$ and to
display avoided-level crossing at $\protect\varepsilon _{0}=0$ with a
minimal distance $\Delta $. The system is excited when the characteristic
qubit frequency is about a multiple of the driving frequency, $\Delta
E/\hbar =k\protect\omega $. Several possible classical realizations have
been demonstrated: (b)~two weakly coupled spring oscillators~\protect\cite%
{Frimmer14}, (c)~a two-mode nano-beam~\protect\cite{Seitner16},
(d)~optomechanical system with two cantilevers~\protect\cite{Fu16}, (e)~two
coupled electrical resonators~\protect\cite{Garrido02, Muirhead16}, (f)~two
coupled polarization modes in an optical ring resonator~\protect\cite%
{Spreeuw90}.}
\label{Fig:schemes}
\end{figure}

Classical oscillators are ubiquitous in nature. With some modifications,
they provide analogues of systems from other fields of physics. An important
example considered here is a basic system of quantum mechanics and quantum
technologies: a two-level system, or qubit \cite{Buluta11, You11, Gu17}. A
qubit is described by its tuned two energy levels, as illustrated in Fig.~%
\ref{Fig:schemes}(a). Being driven, such system experiences resonant
transitions, which is important for both system characterization and
control. However, in a number of works in different contexts, it was argued
that diverse classical systems can behave like qubits. Such systems include
mechanical, opto-mechanical, electrical, plasmonic, and optical
realizations, as illustrated in Fig.~\ref{Fig:schemes}(b-f).

Models based on classical oscillators were used to describe such phenomena
as stimulated resonance Raman effect \cite{Hemmer88}, electromagnetically
induced transparency and Autler-Townes splitting \cite{Garrido02, Peng14},
Landau-Zener transitions \cite{Maris88, Shore09, Novotny10, Faust12}, rapid
adiabatic passage \cite{Shore09}, Rabi oscillations \cite{Faust13, Frimmer14}%
, St\"{u}ckelberg interference \cite{Fu16, Seitner16, Seitner17}, Fano
resonances \cite{Joe06}, squeezed states \cite{Mahboob16}, strong coupling
\cite{Rodriguez16}, light-matter interaction \cite{Frimmer17}, and dynamical
localization \cite{Fu17}. In these works, classical resonators displayed
features which are sometimes thought of as fingerprints of quantum coherent
behaviour. Also such formulations should be differentiated from the layouts
where classical mechanical resonators are used to probe coherent phenomena
in quantum systems, like in Ref.~[\citeonline{LaHaye09}] (see also Ref.~[%
\citeonline{Wei06}]), where a classical nanomechanical resonator is used to
probe quantum phenomena in superconducting qubits. Nevertheless, we will not
discuss other possible realizations of analogues between classical and
quantum behaviour, which can also be found for tunneling in Josephson
junctions \cite{Blackburn16, Shevchenko08, Omelyanchouk08} and light
propagation in periodic optical structures \cite{Longhi11, Eichelkraut14}.

The analogy between classical and quantum phenomena is an intriguing and
very broad subject, see, e.g., Refs.~%
\citeonline{Dragoman, Lambert10, Bliokh14,
Emary14, Miranowicz15, Miranowicz15b}. However, we would like to limit this
work to only classical analogues of strongly-driven qubits. Without taking
dissipation into account, a qubit is described by the Schr\"{o}dinger
equation, which (for the pseudospin 1/2) is equivalent to the classical
Landau-Lifshitz equation \cite{Garanin04}. The extension of this formalism
in the presence of dissipation is known as the Landau-Lifshitz-Bloch
equation \cite{Garanin97, Wieser16, Klenov17}. In this way, any classical
system which obeys the Landau-Lifshitz equation can mimic a qubit's
behaviour.

Besides being fundamentally interesting, such approach of finding
classical-quantum analogies is attractive for the classical analogue
simulation of quantum systems (e.g.,~[\citeonline{Rahimi16}]). In this paper
we continue the work in this direction and address other phenomena such as
Landau-Zener-St\"{u}ckelberg-Majorana (LZSM) interferometry \cite%
{Shevchenko10, Chatterjee18}, latching modulation \cite{Silveri15, Ono18},
and motional averaging \cite{Li13}.

In the rest of this paper, we first present details of how the description
of two classical oscillators can be reduced to the qubit equations of
motion. We further consider situations when the mechanical-resonator system
is driven by strong periodic or noise signals. We demonstrate the resulting
interference fringes, which are remarkably similar to those in the quantum
analogues.

\section{Schr\"{o}dinger-like classical equation of motion}

To be specific, among diverse classical analogues of qubits, we consider
mechanical layouts. Such oscillators can be realized as separate resonators
\cite{Okamoto13, Deng16, Yamaguchi17} or as two modes of one resonator \cite%
{Faust12, Faust13}. Among the advantages of such mechanical systems are that
they operate at room temperature, have small size and high quality factors,
and are only weakly coupled to the environment. Such system is described by
the equations of motion%
\begin{equation}
m\overset{\cdot \cdot }{u}_{i}+m\gamma \dot{u}_{i}+k_{i}u_{i}+k_{\mathrm{c}%
}(u_{i}-u_{j})=0,  \label{eq_for_u_n}
\end{equation}%
where the displacement $u_{i}$ relates to the $i$-th oscillator ($i=1,2$, $%
j\neq i$). For oscillators we assume equal effective masses $m$, equal
damping rates $\gamma $, different spring constants $k_{i}=m\omega _{i}^{2}$%
, and weak coupling, $k_{\mathrm{c}}\ll k_{i}$, see Fig.~\ref{Fig:schemes}%
(b).

It is important to stress that the other layouts in Fig.~\ref{Fig:schemes}
are also described by the same equation (\ref{eq_for_u_n}). Therefore, what
is written below is equally applicable to any classical two-state system.
For example, for two coupled electrical resonators, Fig.~\ref{Fig:schemes}%
(e), instead of displacement we have charges on respective capacitances, $%
u_{i}\rightarrow q_{i}$, the role of masses is played by inductances, $%
m\rightarrow L$, capacitances replace the spring constants, $%
k_{i}(t)\rightarrow C_{i}^{-1}(t)$ and $k_{\mathrm{c}}\rightarrow C_{\mathrm{%
c}}$, and the damping is defined by the resistances, $\gamma \rightarrow R/L$%
. Thus, the present scheme, inspired by Ref.~[\citeonline{Garrido02}],
requires driving via capacitances. Alternatively, the scheme may be modified
so that to be driven via inductances, as in Ref.~[\citeonline{Muirhead16}].
In addition, Fig.~\ref{Fig:schemes}(d) presents two cantilevers, one of
which is coupled to the optical cavity \cite{Fu16}; and in Fig.~\ref%
{Fig:schemes}(f) the two polarization modes of the light propagating in
counter-clockwise (ccw) direction are shown to be tuned by the electro-optic
modulators, EOM1 and EOM2, with the tuning parameter being the electric
field inside the EOM1 \cite{Spreeuw90}.

To obtain the analogue of a driven qubit, following Refs.~[%
\citeonline{Novotny10, Frimmer14}], we assume%
\begin{equation}
k_{1,2}=k_{0}\pm \Delta k(t).  \label{k12}
\end{equation}%
Here $\Delta k(t)$ contains, in general, both dc and ac components. Then,
introducing the interaction-shifted frequency%
\begin{equation}
\Omega _{0}^{2}=\frac{k_{0}+k_{\mathrm{c}}}{m},  \label{W0}
\end{equation}%
the two equations (\ref{eq_for_u_n}) can be rewritten in the matrix form
using the notation of the Pauli matrices $\sigma _{x,z}$:%
\begin{equation}
\left( \frac{d^{2}}{dt^{2}}+\gamma \frac{d}{dt}+\Omega _{0}^{2}\right) \left[
\begin{array}{c}
u_{1} \\
u_{2}%
\end{array}%
\right] -\left( \frac{k_{\mathrm{c}}}{m}\sigma _{x}+\frac{\Delta k}{m}\sigma
_{z}\right) \left[
\begin{array}{c}
u_{1} \\
u_{2}%
\end{array}%
\right] =0.  \label{Eq4}
\end{equation}%
Using the ansatz%
\begin{equation}
\widetilde{u}_{i}=\psi _{i}\exp \left( i\Omega _{0}t\right) ,\text{ \ \ }%
u_{i}=\text{Re }\widetilde{u}_{i},  \label{ansatz}
\end{equation}%
we obtain the equation%
\begin{equation}
\left( \frac{d^{2}}{dt^{2}}+\left( \gamma +2i\Omega _{0}\right) \frac{d}{dt}%
+i\Omega _{0}\gamma \right) \left(
\begin{array}{c}
\psi _{1} \\
\psi _{2}%
\end{array}%
\right) -\left( \frac{k_{\mathrm{c}}}{m}\sigma _{x}+\frac{\Delta k(t)}{m}%
\sigma _{z}\right) \left(
\begin{array}{c}
\psi _{1} \\
\psi _{2}%
\end{array}%
\right) =0.  \label{with_d2/dt2}
\end{equation}

Instead of first directly solving these classical equations of motion for
specific realizations (as e.g. in Refs.~[%
\citeonline{Garrido02, Joe06,
Faust12, Muirhead16}]), we rather demonstrate the equivalence of these to
the motion equation of a qubit, and then (in the next section) we will make
use of the available solutions. We find this appropriate and pedagogical to
first demonstrate the equivalence and then make use of the techniques and
solutions available for qubit dynamics. Our approach differs only in details
from the other authors' methods, though. For example, in Ref.~[%
\citeonline{Frimmer14}] the eigenfrequencies are found and only then the
Bloch-like equation is obtained, while we do vice versa.

Equation (\ref{with_d2/dt2}) is simplified as follows. First, due to small
dissipation, $\gamma $ is neglected next to $\Omega _{0}$. Then, the
slowly-varying envelope approximation \cite{Shore09, Frimmer14} allows
neglecting the second derivative (cf. Ref.~[\citeonline{Okamoto13}]). This
means that $\psi _{n}$ changes little during the time $2\pi /\Omega _{0}$,
i.e. its characteristic evolution frequency $\omega $ is much smaller than $%
\Omega _{0}$, $\omega \ll \Omega _{0}$. Then introducing new notations, from
Eq.~(\ref{with_d2/dt2}) we obtain
\begin{subequations}
\begin{eqnarray}
i\frac{d}{dt}\left\vert \psi \right\rangle &=&H(t)\left\vert \psi
\right\rangle -i\frac{\gamma }{2}\left\vert \psi \right\rangle ,
\label{Schr-like} \\
\left\vert \psi \right\rangle &=&\left(
\begin{array}{c}
\psi _{1} \\
\psi _{2}%
\end{array}%
\right) ,  \label{psi} \\
H(t) &=&\frac{\Delta }{2}\sigma _{x}+\frac{\varepsilon (t)}{2}\sigma _{z},
\label{H} \\
\Delta &=&\frac{k_{\mathrm{c}}}{m\Omega _{0}}\text{ \ }\approx \text{ \ }%
\frac{k_{\mathrm{c}}}{\sqrt{mk_{0}}},  \label{D} \\
\varepsilon (t) &=&\frac{\Delta k(t)}{m\Omega _{0}}\text{ \ }\approx \text{
\ }\frac{\Delta k(t)}{\sqrt{mk_{0}}}.  \label{e}
\end{eqnarray}%
In the absence of dissipation, $\gamma =0$, equation (\ref{Schr-like})
formally coincides with the Schr\"{o}dinger equation for a two-level system
with the Hamiltonian $H(t)$, applying $\hbar =1$.

Dissipation can be eliminated from the problem by the substitution $%
\left\vert \psi \right\rangle =\left\vert \overline{\psi }\right\rangle \exp
\left( -\frac{\gamma }{2}t\right) $, then Eq.~(\ref{Schr-like}) becomes
\end{subequations}
\begin{equation}
i\frac{d}{dt}\left\vert \overline{\psi }\right\rangle =H(t)\left\vert
\overline{\psi }\right\rangle .
\end{equation}%
Alternatively, the \textquotedblleft density matrix\textquotedblright\ can
be introduced as $\rho =\left\vert \psi \right\rangle \left\langle \psi
\right\vert $, where $\left\langle \psi \right\vert :=\left( \psi _{1}^{\ast
},\psi _{2}^{\ast }\right) $. Then for the derivative we obtain%
\begin{equation}
\dot{\rho}=-i\left[ H,\rho \right] -\gamma \rho .  \label{Bloch}
\end{equation}%
This coincides with the Bloch equation for a two-level system with the
Hamiltonian $H(t)$, assuming $\hbar =1$, and with equal relaxation rates, $%
T_{1}=T_{2}=1/\gamma $.

Introducing a convenient parametrization for the \textquotedblleft
Hamiltonian\textquotedblright\ and the \textquotedblleft density
matrix\textquotedblright ,%

\begin{equation}
H(t) = \frac{\Delta }{2}\sigma _{x}+\frac{\varepsilon (t)}{2}\sigma
_{z}\equiv \frac{1}{2}\mathbf{B}\boldsymbol{\sigma },
\end{equation}%

\begin{equation}
\rho  = 1+X\sigma _{x}+Y\sigma _{y}+Z\sigma _{z}\equiv 1+\mathbf{X}%
\boldsymbol{\sigma },  \nonumber
\end{equation}%

the \textquotedblleft Bloch\textquotedblright\ equation (\ref{Bloch}) can be
rewritten in the form of the Landau-Lifshitz-Bloch equation:%
\begin{equation}
\frac{d}{dt}\mathbf{X}=\mathbf{B}\times \mathbf{X}-\gamma \mathbf{X},
\label{Bloch-type}
\end{equation}%
where
\[
\boldsymbol{\sigma }=(\sigma _{x},\sigma _{y},\sigma _{z}),\text{ \ \ }%
\mathbf{X}=\left( X,Y,Z\right) ,\text{ \ \ }\mathbf{B}=\left( \Delta
,0,\varepsilon (t)\right) .
\]%
The diagonal components of the density matrix $\rho $ and the $Z$-component
of the Bloch vector $\mathbf{X}$ define the occupation of the respective
states, while the off-diagonal components of $\rho $ and $X$ and $Y$
describe the coherence.

We now see the analogy between the classical system and a qubit, described
by the Bloch equation. This allows one to expect very similar dynamical
phenomena. This was described in the introduction, while specific results
are presented in the next two sections. After discussing this analogy, let
us point out three key issues (see also, e.~g.,~[%
\citeonline{Frimmer14,
Faust13}]).

First, instead of different energy and phase relaxation rates for a generic
quantum two-level system, for a classical analogue they coincide: $%
T_{1}^{-1}=T_{2}^{-1}=\gamma $. Recall that we have considered identical oscillators, with equal $m$, $k_{0}$, and $\gamma $.  In general, all of these quantities should be different.  Thus, the equivalence drawn between $T_{1}$ and $T_{2}$ for the classical system so far is not general. They are only equivalent here because it is assumed that the damping for both oscillators is the same, which in general will not be the case, particularly for different frequency mechanical oscillators. In general, $T_{1}$ and $T_{2}$ are not related for classical systems, as they would be for a quantum system.  For example, $1/T_{2}=1/2 T_{1}+1/T_{\phi}$ for a quantum system, but not in the case of two classical coupled oscillators. For two oscillators with only damping rates different, $\gamma_{1,2}$, we would have $T_{2}^{-1}=(\gamma_{1}+\gamma_{2})/2$.

Second, a qubit at zero temperature relaxes to the ground state, defined by
the Bloch vector $\mathbf{X}=\left( 0,0,1\right) $, while the classical
analogue in equilibrium relaxes to the zero Bloch-type vector $\mathbf{X}%
=\left( 0,0,0\right) $, as can be seen from Eq.~(\ref{Bloch-type}). This
difference originates from the absence of a classical analogue to the purely
quantum process of quantum emission. \cite{Frimmer14}

Third, one should remember the approximations done: when neglecting the
second time derivative we assumed that the classical system's characteristic
evolution frequency $\omega $ is much smaller than $\Omega _{0}$, i.~e. $%
\omega \ll \Omega _{0}$.

We note that the specific parameters depend on the choice of the system, as
it was outlined in the introduction. To show specific numbers, we can take
the ones close to Ref.~[\citeonline{Faust12}] for a nanomechanical two-mode
beam: $m\simeq 10^{-15}~$g, $k_{0}\simeq 3~$N/m, $k_{\mathrm{c}}\simeq
0.003~ $N/m $\ll k_{0}$, $\gamma \simeq 80$ Hz$\cdot 2\pi $. These
parameters give the following: $\Omega _{0}\simeq 6$ MHz$\cdot 2\pi $, which
indeed satisfies $\Omega _{0}\gg \gamma $, then, $\Delta $ $\simeq 7$ kHz$%
\cdot 2\pi $, which makes driving with $\omega \sim \Delta $ feasible, and
also $\Delta k\sim 0.03~$N/m, which allows to discuss the regime of strong
driving, with the amplitude $A\gg \omega $, $\Delta $.

\section{Solutions of the Schr\"{o}dinger-like equation}

Using the analogy between the dynamical equations for the two coupled
mechanical resonators and a two-level system, one can rewrite results from
the respective publications, e.g. from Refs.~[%
\citeonline{Ashhab07,
Shevchenko10, Shevchenko12}]. For convenience, some analytical results are
written down below, while detailed solutions are illustrated in the next
Section.

For the stationary Hamiltonian (\ref{H}) with the time-independent bias $%
\varepsilon =\varepsilon _{0}$, we can transform from the functions $\psi
_{i}$ to $\psi _{\pm }$, which define the eigenstates, analogously to the
diagonalization of the Hamiltonian for qubits, e.g.\ Ref.~[%
\citeonline{Shevchenko10}]:%
\begin{equation}
i\frac{d}{dt}\left(
\begin{array}{c}
\psi _{-} \\
\psi _{+}%
\end{array}%
\right) =-\frac{\omega _{0}}{2}\sigma _{z}\left(
\begin{array}{c}
\psi _{-} \\
\psi _{+}%
\end{array}%
\right) ,
\end{equation}

\[
\omega _{0}=\sqrt{\Delta ^{2}+\varepsilon _{0}^{2}}.
\]%
The solution of these two uncoupled equations is
\[
\psi _{\pm }(t)=\psi _{\pm }(0)\exp \left( \mp i\frac{\omega _{0}}{2}%
t\right) .
\]%
Together with the ansatz~(\ref{ansatz}), this gives eigenfrequencies for the
oscillations:
\begin{equation}
\Omega _{\pm }=\Omega _{0}\mp \frac{\omega _{0}}{2}=\Omega _{0}\mp \frac{1}{2%
}\sqrt{\Delta ^{2}+\varepsilon _{0}^{2}},  \label{eigenfreq}
\end{equation}%
which display the avoided-level crossing at the zero offset $\varepsilon
_{0}=0$. (This coincides with the eigenfrequencies from Ref.~[%
\citeonline{Frimmer14}], assuming $\Delta \omega \ll $ $\Omega _{0}$, i.e. $%
\Omega _{d},\Omega _{c}\ll \Omega _{0}$ therein.)

The eigen-functions $\psi _{\pm }$ are the amplitudes of the respective
eigen-modes. These are analogous to the energy-level occupation amplitudes
for qubits. Accordingly, we will be interested in the \textquotedblleft
occupation\textquotedblright\ of one of the modes, namely, $\left\vert \psi
_{+}\right\vert ^{2}$, which can be related to a qubit upper-level
occupation probability. In experiments such value can be probed as an
amplitude of the oscillations of the respective mode.\cite{Faust12} For
example, if initially one, say \textquotedblleft $-$\textquotedblright ,
mode is excited, the problem can be formulated in finding the amplitude of
the other mode; in this sense, the value $\left\vert \psi _{+}\right\vert
^{2}$ can be interpreted as a transition probability, describing the
transition from one mode to another.

Consider now several regimes for the dc + ac driving:%
\begin{equation}
\varepsilon (t)=\varepsilon _{0}+A\sin \omega t.  \label{harmonic}
\end{equation}

For the single passage of the avoided crossing, we have the Landau-Zener
problem, for which the solution is given by the probability
\begin{eqnarray}
\left\vert \psi _{+}\right\vert ^{2} &=&P_{\mathrm{LZ}}\ =\ \exp (-2\pi
\delta ), \\
\delta &=&\frac{\Delta ^{2}}{4v},  \nonumber \\
v &=&A\omega \sqrt{1-\left( \varepsilon _{0}/A\right) ^{2}}.  \nonumber
\end{eqnarray}%
This coincides with the result for the LZ-problem in Refs.~[%
\citeonline{Novotny10, Faust12}] with the linear driving: $v=\alpha =A\omega
$.

Analogously, for the double-passage problem, we have the St\"{u}ckelberg
oscillations:%
\begin{eqnarray}
\left\vert \psi _{+}\right\vert ^{2} &=&4P_{\mathrm{LZ}}\left( 1-P_{\mathrm{%
LZ}}\right) \sin ^{2}\Phi _{\mathrm{St}},  \label{Stuck} \\
\Phi _{\mathrm{St}} &=&\frac{1}{2}\int\limits_{t_{1}}^{t_{2}}\sqrt{\Delta
^{2}+\varepsilon (t)^{2}}\,dt+\widetilde{\varphi }_{\mathrm{S}},  \nonumber
\end{eqnarray}%
where $\widetilde{\varphi }_{\mathrm{S}}=\widetilde{\varphi }_{\mathrm{S}%
}(\delta )$ is a parameter varying from $-\pi /4$, for $\delta \ll 1$,\ to $%
-\pi /2$,$\ $for $\delta \gg 1$ \cite{Seitner17}. We note that the integral
above can be estimated, first, by neglecting $\Delta $, and, second, for $%
\varepsilon _{0}=0$, then the integral is given by a special function: a
full elliptic integral of the second kind \cite{Shevchenko06}.

For the multiple-passage problem, consider here only the fast-passage (small
$\Delta $) limit. Then multi-photon Rabi oscillations are envisaged; next to
the $k$-th resonance, where $\varepsilon _{0}\sim k\omega $, we have%
\begin{eqnarray}
\left\vert \psi _{+}^{(k)}(t)\right\vert ^{2} &=&\frac{1}{2}\frac{\Delta
_{k}^{2}}{\Omega _{\mathrm{R}}^{(k)2}}\left( 1-\cos \Omega _{\mathrm{R}%
}^{(k)}t\right) ,  \label{k-Rabi} \\
\Delta _{k} &=&\Delta J_{k}(A/\omega ),  \nonumber \\
\Omega _{\mathrm{R}}^{(k)} &=&\sqrt{\Delta _{k}^{2}+\left( k\omega
-\left\vert \varepsilon _{0}\right\vert \right) ^{2}}\text{.}  \nonumber
\end{eqnarray}%
For large arguments, the Bessel function has the oscillating asymptote $%
J_{k}(x)\approx \sqrt{2/\pi x}\;\cos \left[ x-\left( 2k+1\right) \pi /4%
\right] $. If necessary, the damping is described by adding the factor $\exp
\left( -\gamma t\right) $ in Eqs.~(\ref{Stuck}, \ref{k-Rabi}). The
time-averaged probability distribution is given by the series of Lorentzians:%
\begin{equation}
\overline{\left\vert \psi _{+}\right\vert ^{2}}=\sum\limits_{k}\frac{\Delta
_{k}^{2}}{\Delta _{k}^{2}+\left( k\omega -\left\vert \varepsilon
_{0}\right\vert \right) ^{2}+\gamma ^{2}}.  \label{multi}
\end{equation}%
This, when plotted as a function of $\varepsilon _{0}$ and $A$, could serve
as a visualization of the LZSM interference, which is the main subject of
the next Section.

An analysis of the above equations allows for interpretations of specific
systems. In our example of two coupled classical oscillators, we have for
the Rabi frequency:
\begin{equation}
\Omega _{\mathrm{R}}\ =\ \Omega _{\mathrm{R}}^{(1)}\ \sim \ \Delta \ \propto
\ k_{\mathrm{c}},
\end{equation}%
which means that the energy transfer between the two oscillators, or between
the two modes of a mechanical resonator, appears with a rate proportional to
the coupling strength \cite{Hemmer88}.

\section{Dynamics and interference in the classical two-state system}

\begin{figure}[t]
\centering{\includegraphics[width=0.55\columnwidth]{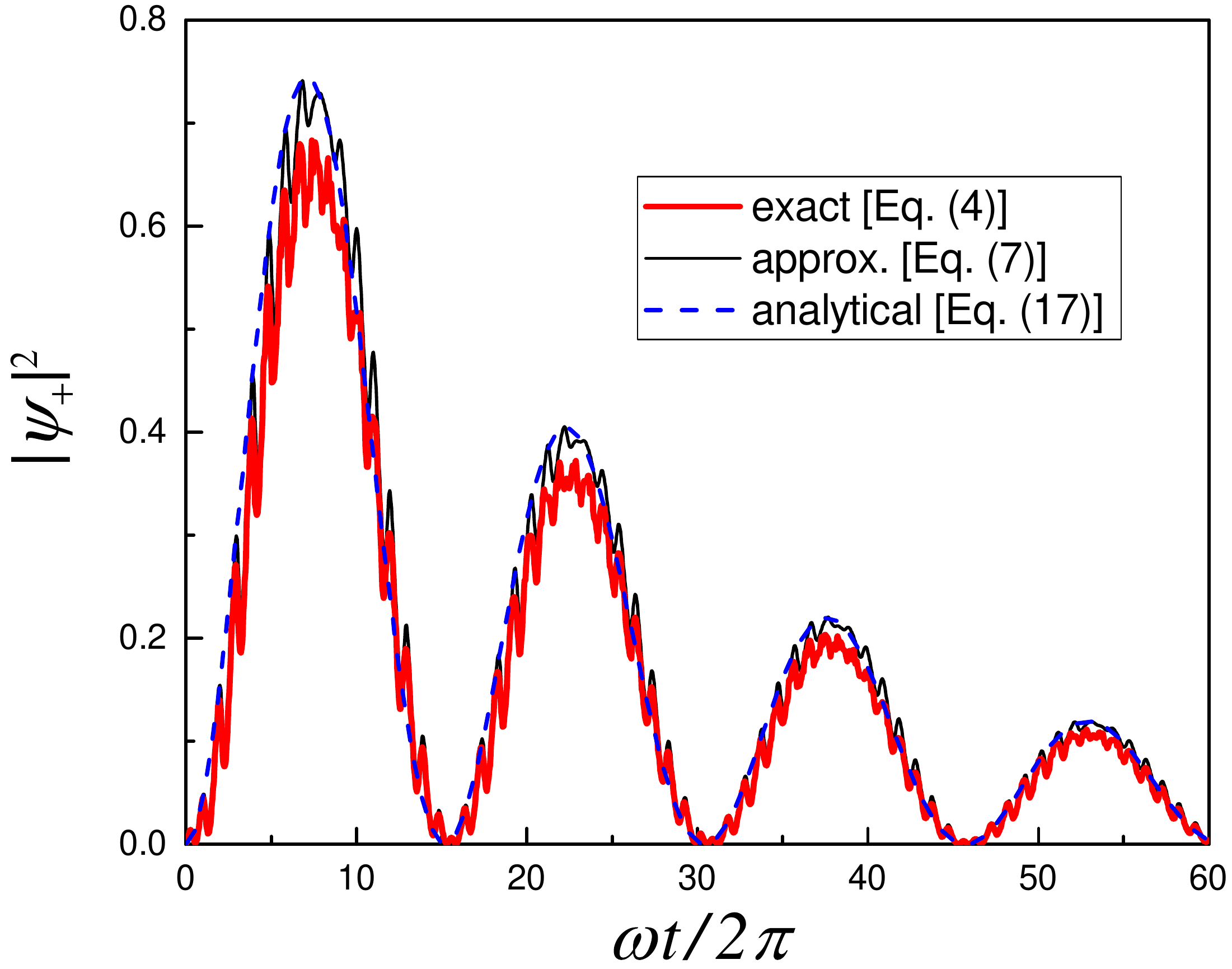}}
\caption{Classical Rabi-like oscillations. When a two-state system is driven
by a resonant signal, with $\protect\omega =\protect\omega _{0}$, the mode
occupation $\left\vert \protect\psi _{+}\right\vert ^{2}$ displays damping
oscillations with Rabi frequency $\Omega _{\mathrm{R}}$. The three curves
display the solutions of the exact equations~(4), the approximate ones (7),
and the analytical solution, Eq.~(17).}
\label{Fig:Rabi}
\end{figure}

The Schr\"{o}dinger-type and Bloch-type equations were presented above for
two coupled mechanical resonators. This was done with several assumptions,
which allowed reducing the original Eq.~(\ref{Eq4}) to Eq.~(\ref{Schr-like}%
). From the latter, a qubit-like behaviour follows. In this section we
confirm and demonstrate this, by solving numerically the original equations
(4). We consider a driven system with the bias $\varepsilon (t)=\varepsilon
_{0}+\varepsilon _{1}(t)$, where $\varepsilon _{1}(t)$ is one of the
following:

(1)\ the sinusoidal function $\varepsilon _{1}(t)=A\sin \omega t$ with weak
or strong amplitudes, so-called Rabi and LZSM regimes, respectively;

(2)\ the rectangular driving with $\varepsilon _{1}(t)=A\,\,\mathrm{sgn}%
(\sin \omega t)$, so-called latching modulation; and

(3)\ a noisy signal, corresponding to random jumps between $\varepsilon
_{1}=+A$ and $-A$, with the characteristic switching frequency $\chi $.

\subsection{Rabi oscillations}

Let us first consider the Rabi regime with a weak-amplitude sinusoidal
driving. For calculations, we here choose the resonant frequency, $\omega
=\omega _{0}$ with $\varepsilon _{0}=5\Delta $, weak amplitude $A=0.7\omega $%
, and significant relaxation $\gamma =0.006\omega $, in order to see the
damping of the Rabi oscillations. In Fig.~\ref{Fig:Rabi}, the thick red
curve shows the numerical solution of the exact Eq.~(\ref{with_d2/dt2}), the
thin black curve is for the numerical solution of the approximate Schr\"{o}%
dinger-like Eq.~(\ref{Schr-like}), and the dashed blue curve depicts the
analytical solution, Eq.~(\ref{k-Rabi}). Similarly to their quantum
counterparts, classical oscillations appear at the same conditions, of weak
resonant driving, and have a similar expression for the Rabi frequency,
Eq.~(17). An important distinction is that the oscillations relax to zero,
in contrast to the quantum case, where resonant oscillations result in a
steady state with nonzero population of the excited state. As we can
conclude from Fig.~\ref{Fig:Rabi}, for non-trivial results we should average
before the system relaxes. This means that the analogue simulation of the
quantum system has to be realized as the dynamics of the classical system on
time scales $\Delta t<\gamma ^{-1}$. In particular, for the resonant
excitation, like the one in Fig.~\ref{Fig:Rabi}, after averaging the
oscillating dynamics, we obtain $\overline{\left\vert \psi _{+}\right\vert
^{2}}\sim 0.5$ for $\gamma \Delta t\ll 1$, while this occupation decreases
for increasing $\gamma \Delta t$.

\subsection{LZSM interferometry}

\begin{figure}[t]
\centering{\includegraphics[width=0.94\columnwidth]{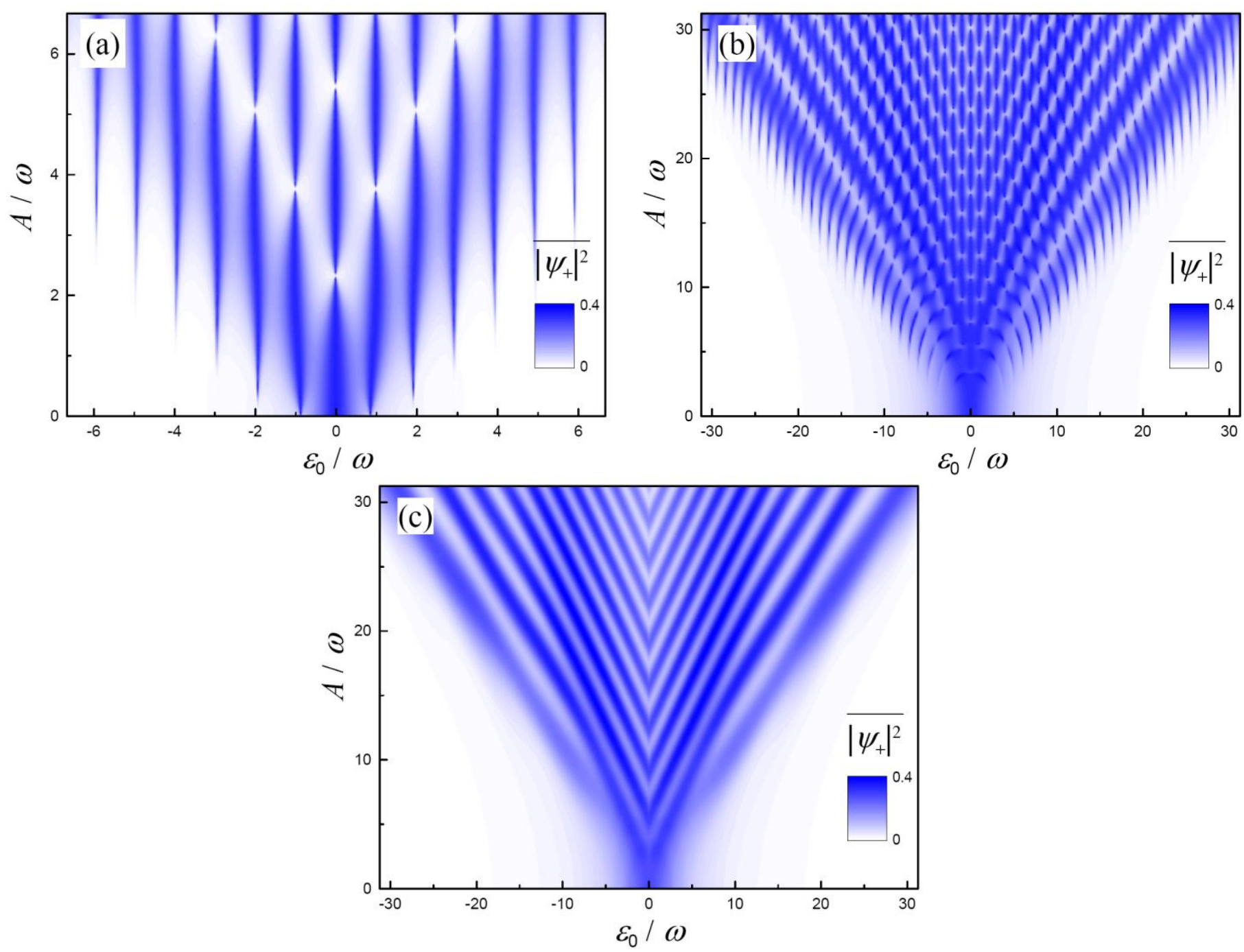} }
\caption{LZSM interferogram for two \textit{classical} oscillators. In (a)
and (b) we present the fast- and slow-driving cases with $\protect\omega %
/\Delta =2$ and $1/3$, respectively, for \textit{small} damping, while panel
(c) demonstrates the case of \textit{stronger} damping for $\protect\omega %
/\Delta =1/3$; see main text for details. Note that analogous LZSM
interferograms were studied for qubits in Ref.~[\noindent
\citeonline{Shevchenko10}]}
\label{Fig:LZSM}
\end{figure}

This regime occurs when a two-level system is strongly driven by a
sinusoidal signal and monitored when changing its parameters; see Ref.~[%
\citeonline{Shevchenko10}] and references therein. In such processes one is
interested in the excitation probability, which increases periodically due
to LZSM transitions between the states. After time averaging, the excitation
probability displays increased values in the vicinity of the multi-photon
resonances, where the energy difference between the states is matched by
multiples of the driving frequency. Changing the system parameters, one can
plot such interference fringes. This approach is useful both for controlling
the system state and for defining the system parameters and its coupling to
the environment \cite{Shevchenko12, Forster14}.

To demonstrate this regime, we now consider the classical-oscillators system
in the strong-driving regime, where, for a driven qubit in the same regime,
LZSM interference takes place. Figure~\ref{Fig:LZSM} shows the time-averaged
mode occupation $\overline{\left\vert \psi _{+}\right\vert ^{2}}$ as a
function of the bias $\varepsilon _{0}$ and the driving amplitude $A$. With
the numerical solution of the exact Eq.~(\ref{with_d2/dt2}) for relatively
high and low frequencies ($\omega \gtrless \Delta $), we obtain the
interferograms in Fig.~\ref{Fig:LZSM}(a) and~(b), respectively. We can
observe an important feature in these graphs: they display that the
excitations appear at the position $\left\vert \varepsilon _{0}\right\vert
=k\omega $, with an integer $k$, analogously to the multi-photon transitions
for qubits, as illustrated by the arrows in Fig.~\ref{Fig:schemes}(a). These
correspond to the minima of the denominator in Eq.~(\ref{multi}), while the
narrowing of these resonance lines appears around the zeros of the Bessel
functions, entering in the numerators in Eq.\ (\ref{multi}). Note that this
results in the interruptions of the resonance lines, where there are no
transitions, even though the resonance condition, $\left\vert \varepsilon
_{0}\right\vert =k\omega $, is fulfilled. This phenomena for a quantum
system is known as the coherent destruction of tunneling.\cite{Grossmann91,
Grifoni98, Miao16}

Note that the interferograms in Fig.~\ref{Fig:LZSM}(a) and~(b) are analogous
to the ones obtained for diverse qubit systems, see Ref.~[%
\citeonline{Shevchenko10}] and references therein. In particular, the
interferograms in Fig.~\ref{Fig:LZSM}(a) and~(b) are analogous to Fig.~7(b)
and Fig.~8(b) in Ref.~[\citeonline{Shevchenko10}], respectively. See also
Ref.~[\citeonline{Parafilo18}] where an analogous interferogram was recently
calculated for a quantum-dot-electromechanical device.

Furthermore, Fig.~\ref{Fig:LZSM}(c) demonstrates the case of relatively
stronger damping, when the interference fringes appear mostly due to the
neighboring two transitions of the avoided crossing in Fig.~\ref{Fig:schemes}%
(a). In this case, the resonances in Fig.~\ref{Fig:LZSM}(c) form
characteristic V-shaped lines. We can observe that the outer lines are
inclined along $A=\pm \varepsilon _{0}$, so that there is no excitation
beyond these lines. This is because at small driving amplitudes, $%
A<\left\vert \varepsilon _{0}\right\vert $, the avoided crossing in Fig.~\ref%
{Fig:schemes}(a) is not reached, which explains the absence of the
excitation. This regime of relatively strong damping with the V-shaped
fringes was called quasiclassical in Ref.~[\citeonline{Berns06}], which was
studied for superconducting and semiconducting qubits in Refs.~[%
\citeonline{Berns06,
Chatterjee18}].

In Fig.~\ref{Fig:LZSM} we have chosen the following parameters for
calculations: relatively weak damping in (a) and (b), $\gamma =0.02\cdot
\omega /2\pi $ and $\gamma =0.1\cdot \omega /2\pi $, respectively, and
stronger damping in (c), $\gamma =\omega /2\pi $. The initial occupation was
zero, $\psi _{+}(t=0)=0$, like in Fig.~\ref{Fig:Rabi}, and then we averaged
for the time interval $\Delta t\sim \gamma ^{-1}$. Namely, we took $\Delta
t=50$, $8$, $1\cdot 2\pi /\omega $ for the three panels in Fig.~\ref%
{Fig:LZSM}, respectively. If we choose $\gamma \Delta t\ll 1$ or $\gamma
\Delta t\gg 1$, \ we would obtain similar data as in Fig.~\ref{Fig:LZSM},
but with a maximum amplitude closer to $0.5$ or $0$, respectively. After
discussing this here, for the rest of the interferograms below we will
assume $\gamma \Delta t\ll 1$.

Alternatively, in addition to the above interferograms, one may be
interested in the dependence on the driving frequency $\omega $, as in
Refs.~[\citeonline{Li13, Silveri15}]. We present such diagram in Fig.~\ref%
{Fig:LZSM2}. Here the driving amplitude $A$ is considered constant and it is
used for normalization, which differs from the interferogram in the previous
figure, where the driving amplitude was the variable value.

\begin{figure}[t]
\centering{\includegraphics[width=0.52\columnwidth]{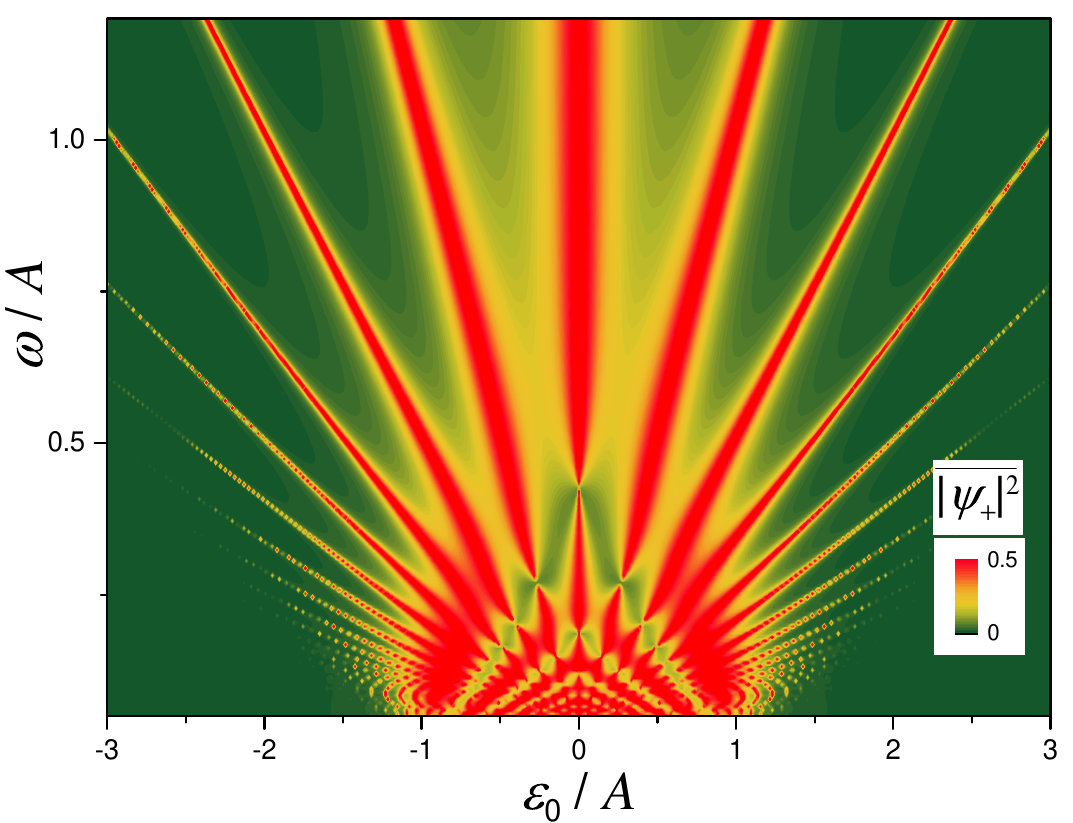}}
\caption{LZSM interferogram showing the dependence of the time-averaged mode
occupation $\overline{\left\vert \protect\psi _{+}\right\vert ^{2}}$ on the
driving frequency $\protect\omega $ and bias $\protect\varepsilon _{0}$.
Note that the analogous interferogram for a qubit was experimentally
observed in~Ref.~[\noindent \citeonline{Silveri15}]. }
\label{Fig:LZSM2}
\end{figure}

\begin{figure}[t]
\centering{\includegraphics[width=0.52\columnwidth]{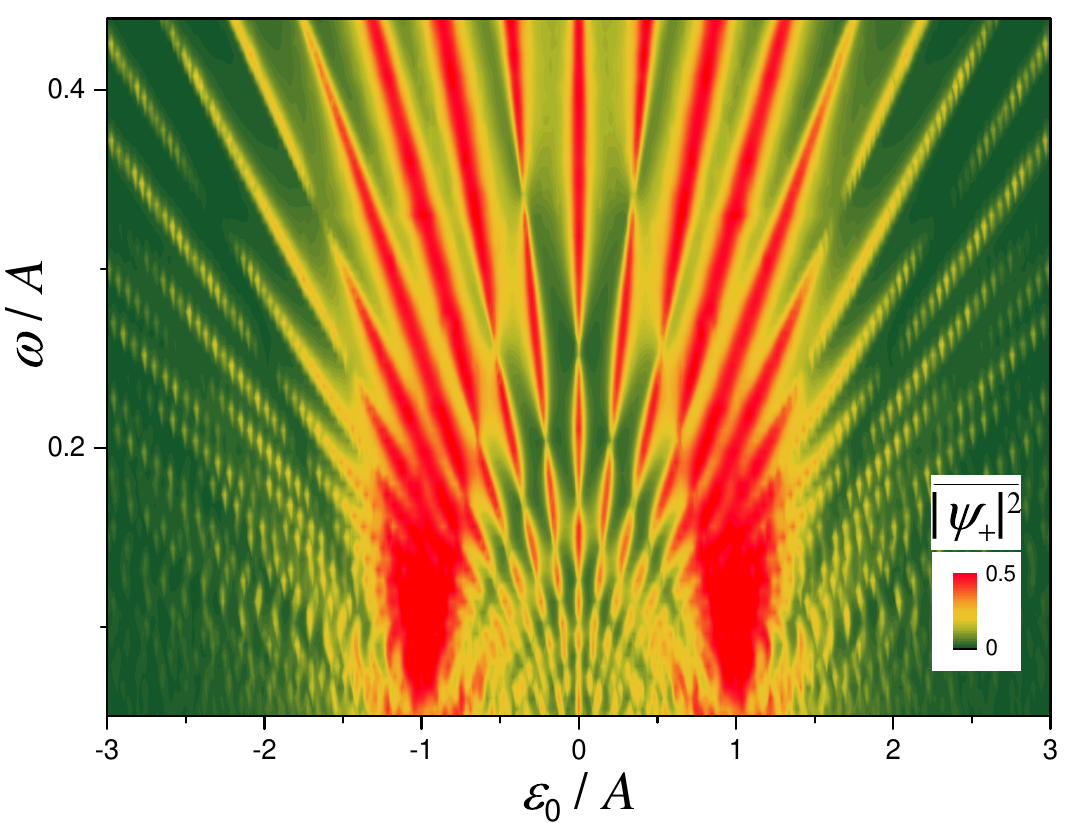}}
\caption{Latching modulation\ with two classical oscillators. The
time-averaged mode occupation $\overline{\left\vert \protect\psi %
_{+}\right\vert ^{2}}$ is plotted when the system is driven by rectangular
pulses. Note that the analogous interferogram for a qubit was experimentally
observed in Ref.~[\noindent \citeonline{Silveri15}].}
\label{Fig:latching}
\end{figure}

\subsection{Latching modulation}

If the system is driven by rectangular pulses instead of a sinusoidal
signal, it experiences periodic fast changes between the two states. Namely,
if the bias is $\varepsilon (t)=\varepsilon _{0}+A\,\,\mathrm{sgn}(\sin
\omega t)$, this means that the system is abruptly latched between the two
limiting states with the bias given by either $\varepsilon _{0}+A$ or $%
\varepsilon _{0}-A$, where it stays in each for about half of the period
\cite{Silveri15, Ono18}. A conceptual distinction from the sinusoidal
driving is in that the avoided-level crossing, where LZSM transitions
appear, is crossed rapidly. That is why the theory developed for smooth
sinusoidal driving had to be revisited for the rectangular-pulse driving,
which was done in Ref.~[\citeonline{Silveri15}] both theoretically and
experimentally for a superconducting qubit. Here, we study such a
formulation for our classical system: two coupled classical oscillators.

To describe this regime, we have solved Eq.~(\ref{with_d2/dt2}), as in the
previous subsection, but now with a different bias. Figure~\ref{Fig:latching}
shows the time-averaged mode occupation $\overline{\left\vert \psi
_{+}\right\vert ^{2}}$ as a function of the bias $\varepsilon _{0}$ and the
driving frequency $\omega $. Such latching modulation displays interference
fringes, different from the ones shown above, for a harmonic driving.
Interestingly, at low frequencies the system is indeed mostly latched to the
two states with the resonances around $\varepsilon _{0}=\pm A$, while for
higher frequencies there is no trace of the latching, and the position of
the resonances is described by the inclined resonance lines. This means
that, due to the interference, our system latching between $\varepsilon
_{0}=\pm A$ displays resonances for any other values of $\varepsilon _{0}$.
Note that here, for the classical system, we obtained a remarkable agreement
with the diagram obtained recently for the experimental qubit in Ref.~[%
\citeonline{Silveri15}].

\subsection{Motional averaging}

\begin{figure}[t]
\centering{\includegraphics[width=0.55\columnwidth]{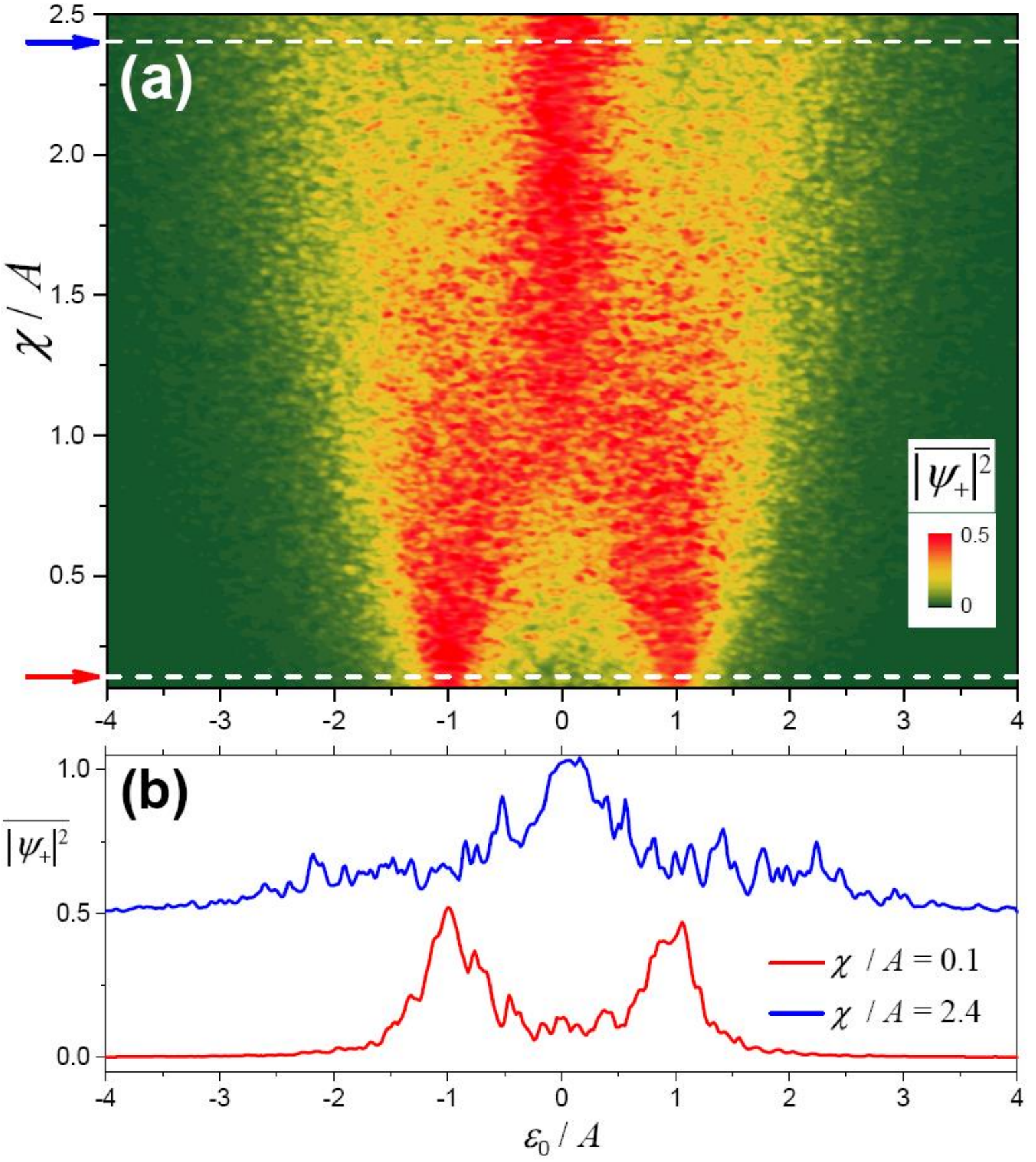} }
\caption{Motional averaging\ with two classical oscillators. (a) The
time-averaged mode occupation $\overline{\left\vert \protect\psi %
_{+}\right\vert ^{2}}$ as a function of the bias $\protect\varepsilon _{0}$
and the jumping rate $\protect\chi $; this displays two peaks at around $%
\protect\varepsilon _{0}=\pm A$ for a slowly jumping signal with $\protect%
\chi \ll A$, while the faster jumping, with $\protect\chi \gtrsim A$,
results in the merging of the two peaks into one. (b) The time-averaged mode
occupation as a function of the bias $\protect\varepsilon _{0}$ for the two
values of the jumping rate $\protect\chi $. Note that analogous dependencies
were experimentally demonstrated for a superconducting qubit in
Ref.~[\noindent \citeonline{Li13}].}
\label{Fig:M-averaging}
\end{figure}

Differently from above, in the regime which we refer to as the motional
averaging, the system is driven by noise rather than by a periodic signal.
The rapid jumps between the two states appear stochastically. For this we
follow Ref.~[\citeonline{Li13}], where it was demonstrated for a qubit that
when increasing the frequency of the jumps, two separate spectral lines
merge into one line. This observation allowed the authors of Ref.~[%
\citeonline{Li13}] to speculate about the analogue simulation of this
so-called motional averaging, originally observed in nuclear magnetic
resonance spectroscopy.

Accordingly, we consider here the classical two-state system driven by a
non-periodic signal. Namely, we take the bias $\varepsilon (t)=\varepsilon
_{0}+\varepsilon _{1}(t)$, admitting random jumps between $\varepsilon
_{1}=+A$ and $-A$. The jumps are assumed to appear with the average jumping
rate $\chi $. In Fig.~\ref{Fig:M-averaging} we plot the time-averaged mode
occupation $\overline{\left\vert \psi _{+}\right\vert ^{2}}$ as a function
of the detuning $\varepsilon _{0}$ and the characteristic switching
frequency $\chi $.

Our data shown in Fig.~\ref{Fig:M-averaging} are consistent with the result
of Ref. [\citeonline{Li13}] in that, at low $\chi $, there are two peaks,
which merge into one, for increasing $\chi $. To make this point explicit,
in Fig.~\ref{Fig:M-averaging}(b) we plot the two cross-sections of Fig.~\ref%
{Fig:M-averaging}(a) along the horizontal white dashed lines. The red curve
in Fig.~\ref{Fig:M-averaging}(b) is plotted for a low switching frequency,
and this displays two distinct peaks at $\varepsilon _{0}=\pm A$. The blue
curve, shifted vertically for clarity, is plotted for a relatively high
switching frequency. This displays a main peak aroung $\varepsilon _{0}=0$,
which can be interpreted as the averaging due to the motion between the two
states.

\section{Conclusions}

Remarkably, the classical system of two weakly coupled classical oscillators
form the doppelg\"{a}nger of a quantum two-level system. Namely, its
equation of motion formally coincides with either the Schr\"{o}dinger
equation or with the Bloch equation in the cases when the relaxation is
either ignored or taken into account, respectively. This means that the
dynamical phenomena of the two-state classical system can be directly
described by the ones already studied for the quantum two-level systems, and
vice versa. This was known and studied some time ago, e.g. in Refs.~[%
\citeonline{Maris88, Hemmer88, Spreeuw90,
Garrido02}]. However, diverse experimental mechanical resonators which are
good enough for analogue simulations appeared only recently \cite{Faust13,
Okamoto13, Deng16, Fu16}. Such mechanical resonators have high quality
factors and have reliable control of their inter-mode coupling. Moreover,
there is recent interest in strongly-driven quantum systems, e.g. Refs.~[%
\citeonline{Shevchenko10, Li13, Silveri15}]. These two
developments stimulated us to further consider the analogy between weakly
and controllably coupled mechanical oscillators and the driven quantum
few-level system. In particular, we demonstrated classical analogues of the
effects recently studied for qubits: Landau-Zener-St\"{u}ckelberg-Majorana
interferometry, latching modulation, and motional averaging. Besides the
pure interest of such dynamical phenomena linking classical and quantum
physics, one may consider simulating some quantum phenomena with classical
systems.

\section*{Acknowledgments}

SNS acknowledges useful discussions with K.~Ono, E.~Weig and H.~Ribeiro. FN
acknowledges useful discussion with S.~Ludwig. This work was partially
supported by the MURI Center for Dynamic Magneto-Optics via the AFOSR Award No.~FA9550-14-1-0040, the Army Research Office (ARO) under grant number 73315PH, the AOARD grant No.~FA2386-18-1-4045, the CREST Grant No.~JPMJCR1676, the IMPACT program of JST, the RIKEN-AIST Challenge Research Fund, the JSPS-RFBR grant No.~17-52-50023, and the Sir John Templeton Foundation, and the State Fund for Fundamental Research of Ukraine (F66/95-2016).

\section*{Author contributions statement}

O.V.I. and S.N.S. did the calculations, F.N. supervised the work. All
authors discussed the results and co-wrote the manuscript.

\section*{Additional information}

\textbf{Competing interests: }The authors declare no competing interests.

\nocite{apsrev41Control}
\bibliographystyle{apsrev4-1}
\bibliography{2NRs_bibliography}

\begin{thebibliography}{10}
\expandafter\ifx\csname url\endcsname\relax
  \def\url#1{\texttt{#1}}\fi
\expandafter\ifx\csname urlprefix\endcsname\relax\def\urlprefix{URL }\fi
\expandafter\ifx\csname doiprefix\endcsname\relax\def\doiprefix{DOI }\fi
\providecommand{\bibinfo}[2]{#2}
\providecommand{\eprint}[2][]{\url{#2}}

\bibitem{Buluta11}
\bibinfo{author}{Buluta, I.}, \bibinfo{author}{Ashhab, S.} \&
  \bibinfo{author}{Nori, F.}
\newblock \bibinfo{journal}{\bibinfo{title}{Natural and artificial atoms for
  quantum computation}}.
\newblock {\emph{\JournalTitle{Rep. Prog. Phys.}}}
  \textbf{\bibinfo{volume}{74}}, \bibinfo{pages}{104401}
  (\bibinfo{year}{2011}).

\bibitem{You11}
\bibinfo{author}{You, J.~Q.} \& \bibinfo{author}{Nori, F.}
\newblock \bibinfo{journal}{\bibinfo{title}{Atomic physics and quantum optics
  using superconducting circuits}}.
\newblock {\emph{\JournalTitle{Nature}}} \textbf{\bibinfo{volume}{474}},
  \bibinfo{pages}{589--597} (\bibinfo{year}{2011}).

\bibitem{Gu17}
\bibinfo{author}{Gu, X.}, \bibinfo{author}{Kockum, A.~F.},
  \bibinfo{author}{Miranowicz, A.}, \bibinfo{author}{Liu, Y.-X.} \&
  \bibinfo{author}{Nori, F.}
\newblock \bibinfo{journal}{\bibinfo{title}{Microwave photonics with
  superconducting quantum circuits}}.
\newblock {\emph{\JournalTitle{Phys. Rep.}}}
  \textbf{\bibinfo{volume}{718-719}}, \bibinfo{pages}{1--102}
  (\bibinfo{year}{2017}).

\bibitem{Hemmer88}
\bibinfo{author}{Hemmer, P.~R.} \& \bibinfo{author}{Prentiss, M.~G.}
\newblock \bibinfo{journal}{\bibinfo{title}{Coupled-pendulum model of the
  stimulated resonance {R}aman effect}}.
\newblock {\emph{\JournalTitle{J. Opt. Soc. Am. B}}}
  \textbf{\bibinfo{volume}{5}}, \bibinfo{pages}{1613} (\bibinfo{year}{1988}).

\bibitem{Garrido02}
\bibinfo{author}{Garrido~Alzar, C.~L.}, \bibinfo{author}{Martinez, M. A.~G.} \&
  \bibinfo{author}{Nussenzveig, P.}
\newblock \bibinfo{journal}{\bibinfo{title}{Classical analog of
  electromagnetically induced transparency}}.
\newblock {\emph{\JournalTitle{Am. J. Phys.}}} \textbf{\bibinfo{volume}{70}},
  \bibinfo{pages}{37} (\bibinfo{year}{2002}).

\bibitem{Peng14}
\bibinfo{author}{Peng, B.}, \bibinfo{author}{\"Ozdemir, S.~K.},
  \bibinfo{author}{Chen, W.}, \bibinfo{author}{Nori, F.} \&
  \bibinfo{author}{Yang, L.}
\newblock \bibinfo{journal}{\bibinfo{title}{What is and what is not
  electromagnetically induced transparency in whispering-gallery
  microcavities}}.
\newblock {\emph{\JournalTitle{Nat. Comm.}}} \textbf{\bibinfo{volume}{5}},
  \bibinfo{pages}{5082} (\bibinfo{year}{2014}).

\bibitem{Maris88}
\bibinfo{author}{Maris, H.~J.} \& \bibinfo{author}{Xiong, Q.}
\newblock \bibinfo{journal}{\bibinfo{title}{Adiabatic and nonadiabatic
  processes in classical and quantum mechanics}}.
\newblock {\emph{\JournalTitle{Am. J. Phys.}}} \textbf{\bibinfo{volume}{56}},
  \bibinfo{pages}{1114--1117} (\bibinfo{year}{1988}).

\bibitem{Shore09}
\bibinfo{author}{Shore, B.~W.}, \bibinfo{author}{Gromovyy, M.~V.},
  \bibinfo{author}{Yatsenko, L.~P.} \& \bibinfo{author}{Romanenko, V.~I.}
\newblock \bibinfo{journal}{\bibinfo{title}{Simple mechanical analogs of rapid
  adiabatic passage in atomic physics}}.
\newblock {\emph{\JournalTitle{Am. J. Phys.}}} \textbf{\bibinfo{volume}{77}},
  \bibinfo{pages}{1183--1194} (\bibinfo{year}{2009}).

\bibitem{Novotny10}
\bibinfo{author}{Novotny, L.}
\newblock \bibinfo{journal}{\bibinfo{title}{Strong coupling, energy splitting,
  and level crossings: A classical perspective}}.
\newblock {\emph{\JournalTitle{Am. J. Phys.}}} \textbf{\bibinfo{volume}{78}},
  \bibinfo{pages}{1199--1202} (\bibinfo{year}{2010}).

\bibitem{Faust12}
\bibinfo{author}{Faust, T.} \emph{et~al.}
\newblock \bibinfo{journal}{\bibinfo{title}{Nonadiabatic dynamics of two
  strongly coupled nanomechanical resonator modes}}.
\newblock {\emph{\JournalTitle{Phys. Rev. Lett.}}}
  \textbf{\bibinfo{volume}{109}}, \bibinfo{pages}{037205}
  (\bibinfo{year}{2012}).

\bibitem{Faust13}
\bibinfo{author}{Faust, T.}, \bibinfo{author}{Rieger, J.},
  \bibinfo{author}{Seitner, M.~J.}, \bibinfo{author}{Kotthaus, J.~P.} \&
  \bibinfo{author}{Weig, E.~M.}
\newblock \bibinfo{journal}{\bibinfo{title}{Coherent control of a classical
  nanomechanical two-level system}}.
\newblock {\emph{\JournalTitle{Nat. Phys.}}} \textbf{\bibinfo{volume}{9}},
  \bibinfo{pages}{485--488} (\bibinfo{year}{2013}).

\bibitem{Frimmer14}
\bibinfo{author}{Frimmer, M.} \& \bibinfo{author}{Novotny, L.}
\newblock \bibinfo{journal}{\bibinfo{title}{The classical {B}loch equations}}.
\newblock {\emph{\JournalTitle{Am. J. Phys.}}} \textbf{\bibinfo{volume}{82}},
  \bibinfo{pages}{947--954} (\bibinfo{year}{2014}).

\bibitem{Fu16}
\bibinfo{author}{Fu, H.} \emph{et~al.}
\newblock \bibinfo{journal}{\bibinfo{title}{Classical analog of {S}t\"uckelberg
  interferometry in a two-coupled-cantilever--based optomechanical system}}.
\newblock {\emph{\JournalTitle{Phys. Rev. A}}} \textbf{\bibinfo{volume}{94}},
  \bibinfo{pages}{043855} (\bibinfo{year}{2016}).

\bibitem{Seitner16}
\bibinfo{author}{Seitner, M.~J.} \emph{et~al.}
\newblock \bibinfo{journal}{\bibinfo{title}{Classical {S}t\"uckelberg
  interferometry of a nanomechanical two-mode system}}.
\newblock {\emph{\JournalTitle{Phys. Rev. B}}} \textbf{\bibinfo{volume}{94}},
  \bibinfo{pages}{245406} (\bibinfo{year}{2016}).

\bibitem{Seitner17}
\bibinfo{author}{Seitner, M.~J.}, \bibinfo{author}{Ribeiro, H.},
  \bibinfo{author}{K\"olbl, J.}, \bibinfo{author}{Faust, T.} \&
  \bibinfo{author}{Weig, E.~M.}
\newblock \bibinfo{journal}{\bibinfo{title}{Finite-time {S}t\"uckelberg
  interferometry with nanomechanical modes}}.
\newblock {\emph{\JournalTitle{New J. Phys.}}} \textbf{\bibinfo{volume}{19}},
  \bibinfo{pages}{033011} (\bibinfo{year}{2017}).

\bibitem{Joe06}
\bibinfo{author}{Joe, Y.~S.}, \bibinfo{author}{Satanin, A.~M.} \&
  \bibinfo{author}{Kim, C.~S.}
\newblock \bibinfo{journal}{\bibinfo{title}{Classical analogy of {F}ano
  resonances}}.
\newblock {\emph{\JournalTitle{Physica Scripta}}}
  \textbf{\bibinfo{volume}{74}}, \bibinfo{pages}{259} (\bibinfo{year}{2006}).

\bibitem{Mahboob16}
\bibinfo{author}{Mahboob, I.}, \bibinfo{author}{Okamoto, H.} \&
  \bibinfo{author}{Yamaguchi, H.}
\newblock \bibinfo{journal}{\bibinfo{title}{Enhanced visibility of two-mode
  thermal squeezed states via degenerate parametric amplification and
  resonance}}.
\newblock {\emph{\JournalTitle{New J. Phys.}}} \textbf{\bibinfo{volume}{18}},
  \bibinfo{pages}{083009} (\bibinfo{year}{2016}).

\bibitem{Rodriguez16}
\bibinfo{author}{Rodriguez, S. R.~K.}
\newblock \bibinfo{journal}{\bibinfo{title}{Classical and quantum distinctions
  between weak and strong coupling}}.
\newblock {\emph{\JournalTitle{Eur. J. Phys.}}} \textbf{\bibinfo{volume}{37}},
  \bibinfo{pages}{025802} (\bibinfo{year}{2016}).

\bibitem{Frimmer17}
\bibinfo{author}{Frimmer, M.} \& \bibinfo{author}{Novotny, L.}
\newblock \emph{\bibinfo{title}{Light-Matter Interactions: A Coupled Oscillator
  Description}}, \bibinfo{pages}{3--14} (\bibinfo{publisher}{Springer
  Netherlands}, \bibinfo{address}{Dordrecht}, \bibinfo{year}{2017,
  arXiv:1604.04367}).

\bibitem{Fu17}
\bibinfo{author}{Fu, H.} \emph{et~al.}
\newblock \bibinfo{journal}{\bibinfo{title}{Classical dynamical localization in
  a strongly driven two-mode mechanical system}}.
\newblock {\emph{\JournalTitle{arXiv:1706.06254}}}  (\bibinfo{year}{2017}).

\bibitem{LaHaye09}
\bibinfo{author}{LaHaye, M.~D.}, \bibinfo{author}{Suh, J.},
  \bibinfo{author}{Echternach, P.~M.}, \bibinfo{author}{Schwab, K.~C.} \&
  \bibinfo{author}{Roukes, M.~L.}
\newblock \bibinfo{journal}{\bibinfo{title}{Nanomechanical measurements of a
  superconducting qubit}}.
\newblock {\emph{\JournalTitle{Nature}}} \textbf{\bibinfo{volume}{459}},
  \bibinfo{pages}{960--964} (\bibinfo{year}{2009}).

\bibitem{Wei06}
\bibinfo{author}{Wei, L.~F.}, \bibinfo{author}{Liu, Y.-X.},
  \bibinfo{author}{Sun, C.~P.} \& \bibinfo{author}{Nori, F.}
\newblock \bibinfo{journal}{\bibinfo{title}{Probing tiny motions of
  nanomechanical resonators: Classical or quantum mechanical?}}
\newblock {\emph{\JournalTitle{Phys. Rev. Lett.}}}
  \textbf{\bibinfo{volume}{97}}, \bibinfo{pages}{237201}
  (\bibinfo{year}{2006}).

\bibitem{Blackburn16}
\bibinfo{author}{Blackburn, J.~A.}, \bibinfo{author}{Cirillo, M.} \&
  \bibinfo{author}{Gronbech-Jensen, N.}
\newblock \bibinfo{journal}{\bibinfo{title}{A survey of classical and quantum
  interpretations of experiments on {J}osephson junctions at very low
  temperatures}}.
\newblock {\emph{\JournalTitle{Phys. Rep.}}} \textbf{\bibinfo{volume}{611}},
  \bibinfo{pages}{1 -- 33} (\bibinfo{year}{2016}).

\bibitem{Shevchenko08}
\bibinfo{author}{Shevchenko, S.~N.}, \bibinfo{author}{Omelyanchouk, A.~N.},
  \bibinfo{author}{Zagoskin, A.~M.}, \bibinfo{author}{Savel'ev, S.} \&
  \bibinfo{author}{Nori, F.}
\newblock \bibinfo{journal}{\bibinfo{title}{Distinguishing quantum from
  classical oscillations in a driven phase qubit}}.
\newblock {\emph{\JournalTitle{New J. Phys.}}} \textbf{\bibinfo{volume}{10}},
  \bibinfo{pages}{073026} (\bibinfo{year}{2008}).

\bibitem{Omelyanchouk08}
\bibinfo{author}{Omelyanchouk, A.~N.}, \bibinfo{author}{Shevchenko, S.~N.},
  \bibinfo{author}{Zagoskin, A.~M.}, \bibinfo{author}{Il'ichev, E.} \&
  \bibinfo{author}{Nori, F.}
\newblock \bibinfo{journal}{\bibinfo{title}{Pseudo-{R}abi oscillations in
  superconducting flux qubits in the classical regime}}.
\newblock {\emph{\JournalTitle{Phys. Rev. B}}} \textbf{\bibinfo{volume}{78}},
  \bibinfo{pages}{054512} (\bibinfo{year}{2008}).

\bibitem{Longhi11}
\bibinfo{author}{Longhi, S.}
\newblock \bibinfo{journal}{\bibinfo{title}{Classical simulation of
  relativistic quantum mechanics in periodic optical structures}}.
\newblock {\emph{\JournalTitle{Appl. Phys. B}}} \textbf{\bibinfo{volume}{104}},
  \bibinfo{pages}{453} (\bibinfo{year}{2011}).

\bibitem{Eichelkraut14}
\bibinfo{author}{Eichelkraut, T.} \emph{et~al.}
\newblock \bibinfo{journal}{\bibinfo{title}{Coherent random walks in free
  space}}.
\newblock {\emph{\JournalTitle{Optica}}} \textbf{\bibinfo{volume}{1}},
  \bibinfo{pages}{268--271} (\bibinfo{year}{2014}).

\bibitem{Dragoman}
\bibinfo{author}{Dragoman, D.} \& \bibinfo{author}{Dragoman, M.}
\newblock \emph{\bibinfo{title}{Quantum-Classical Analogies}}
  (\bibinfo{publisher}{Springer}, \bibinfo{year}{2004}).

\bibitem{Lambert10}
\bibinfo{author}{Lambert, N.}, \bibinfo{author}{Emary, C.},
  \bibinfo{author}{Chen, Y.-N.} \& \bibinfo{author}{Nori, F.}
\newblock \bibinfo{journal}{\bibinfo{title}{Distinguishing quantum and
  classical transport through nanostructures}}.
\newblock {\emph{\JournalTitle{Phys. Rev. Lett.}}}
  \textbf{\bibinfo{volume}{105}}, \bibinfo{pages}{176801}
  (\bibinfo{year}{2010}).

\bibitem{Bliokh14}
\bibinfo{author}{Bliokh, K.~Y.}, \bibinfo{author}{Bekshaev, A.~Y.},
  \bibinfo{author}{Kofman, A.~G.} \& \bibinfo{author}{Nori, F.}
\newblock \bibinfo{journal}{\bibinfo{title}{Photon trajectories, anomalous
  velocities and weak measurements: a classical interpretation}}.
\newblock {\emph{\JournalTitle{New J. Phys.}}} \textbf{\bibinfo{volume}{15}},
  \bibinfo{pages}{073022} (\bibinfo{year}{2013}).

\bibitem{Emary14}
\bibinfo{author}{Emary, C.}, \bibinfo{author}{Lambert, N.} \&
  \bibinfo{author}{Nori, F.}
\newblock \bibinfo{journal}{\bibinfo{title}{{Leggett-Garg} inequalities}}.
\newblock {\emph{\JournalTitle{Rep. Prog. Phys.}}}
  \textbf{\bibinfo{volume}{77}}, \bibinfo{pages}{016001}
  (\bibinfo{year}{2014}).

\bibitem{Miranowicz15}
\bibinfo{author}{Miranowicz, A.} \emph{et~al.}
\newblock \bibinfo{journal}{\bibinfo{title}{Statistical mixtures of states can
  be more quantum than their superpositions: Comparison of nonclassicality
  measures for single-qubit states}}.
\newblock {\emph{\JournalTitle{Phys. Rev. A}}} \textbf{\bibinfo{volume}{91}},
  \bibinfo{pages}{042309} (\bibinfo{year}{2015}).

\bibitem{Miranowicz15b}
\bibinfo{author}{Miranowicz, A.}, \bibinfo{author}{Bartkiewicz, K.},
  \bibinfo{author}{Lambert, N.}, \bibinfo{author}{Chen, Y.-N.} \&
  \bibinfo{author}{Nori, F.}
\newblock \bibinfo{journal}{\bibinfo{title}{Increasing relative nonclassicality
  quantified by standard entanglement potentials by dissipation and unbalanced
  beam splitting}}.
\newblock {\emph{\JournalTitle{Phys. Rev. A}}} \textbf{\bibinfo{volume}{92}},
  \bibinfo{pages}{062314} (\bibinfo{year}{2015}).

\bibitem{Garanin04}
\bibinfo{author}{Garanin, D.~A.} \& \bibinfo{author}{Schilling, R.}
\newblock \bibinfo{journal}{\bibinfo{title}{Quantum nonlinear spin switching
  model}}.
\newblock {\emph{\JournalTitle{Phys. Rev. B}}} \textbf{\bibinfo{volume}{69}},
  \bibinfo{pages}{104412} (\bibinfo{year}{2004}).

\bibitem{Garanin97}
\bibinfo{author}{Garanin, D.~A.}
\newblock \bibinfo{journal}{\bibinfo{title}{{Fokker-Planck and
  Landau-Lifshitz-Bloch} equations for classical ferromagnets}}.
\newblock {\emph{\JournalTitle{Phys. Rev. B}}} \textbf{\bibinfo{volume}{55}},
  \bibinfo{pages}{3050--3057} (\bibinfo{year}{1997}).

\bibitem{Wieser16}
\bibinfo{author}{Wieser, R.}
\newblock \bibinfo{journal}{\bibinfo{title}{Derivation of a time dependent
  {Schr\"odinger equation as the quantum mechanical Landau-Lifshitz-B}loch
  equation}}.
\newblock {\emph{\JournalTitle{J. Phys.: Cond. Mat.}}}
  \textbf{\bibinfo{volume}{28}}, \bibinfo{pages}{396003}
  (\bibinfo{year}{2016}).

\bibitem{Klenov17}
\bibinfo{author}{Klenov, N.~V.} \emph{et~al.}
\newblock \bibinfo{journal}{\bibinfo{title}{Flux qubit interaction with rapid
  single-flux quantum logic circuits: Control and readout}}.
\newblock {\emph{\JournalTitle{Low Temp. Phys.}}}
  \textbf{\bibinfo{volume}{43}}, \bibinfo{pages}{789--798}
  (\bibinfo{year}{2017}).

\bibitem{Rahimi16}
\bibinfo{author}{Rahimi-Keshari, S.}, \bibinfo{author}{Ralph, T.~C.} \&
  \bibinfo{author}{Caves, C.~M.}
\newblock \bibinfo{journal}{\bibinfo{title}{Sufficient conditions for efficient
  classical simulation of quantum optics}}.
\newblock {\emph{\JournalTitle{Phys. Rev. X}}} \textbf{\bibinfo{volume}{6}},
  \bibinfo{pages}{021039} (\bibinfo{year}{2016}).

\bibitem{Shevchenko10}
\bibinfo{author}{Shevchenko, S.~N.}, \bibinfo{author}{Ashhab, S.} \&
  \bibinfo{author}{Nori, F.}
\newblock \bibinfo{journal}{\bibinfo{title}{{Landau-Zener-St\"uckelberg}
  interferometry}}.
\newblock {\emph{\JournalTitle{Phys. Rep.}}} \textbf{\bibinfo{volume}{492}},
  \bibinfo{pages}{1 -- 30} (\bibinfo{year}{2010}).

\bibitem{Chatterjee18}
\bibinfo{author}{Chatterjee, A.} \emph{et~al.}
\newblock \bibinfo{journal}{\bibinfo{title}{A silicon-based single-electron
  interferometer coupled to a fermionic sea}}.
\newblock {\emph{\JournalTitle{Phys. Rev. B}}} \textbf{\bibinfo{volume}{97}},
  \bibinfo{pages}{045405} (\bibinfo{year}{2018}).

\bibitem{Silveri15}
\bibinfo{author}{Silveri, M.~P.} \emph{et~al.}
\newblock \bibinfo{journal}{\bibinfo{title}{St\"uckelberg interference in a
  superconducting qubit under periodic latching modulation}}.
\newblock {\emph{\JournalTitle{New J. Phys.}}} \textbf{\bibinfo{volume}{17}},
  \bibinfo{pages}{043058} (\bibinfo{year}{2015}).

\bibitem{Ono18}
\bibinfo{author}{Ono, K.}, \bibinfo{author}{Shevchenko, S.~N.},
  \bibinfo{author}{Mori, T.}, \bibinfo{author}{Moriyama, S.} \&
  \bibinfo{author}{Nori, F.}
\newblock \bibinfo{journal}{\bibinfo{title}{Quantum interferometry with a
  high-temperature single-spin qubit}}.
\newblock {\emph{\JournalTitle{{in preparation}}}} .

\bibitem{Li13}
\bibinfo{author}{Li, J.} \emph{et~al.}
\newblock \bibinfo{journal}{\bibinfo{title}{Motional averaging in a
  superconducting qubit}}.
\newblock {\emph{\JournalTitle{Nat. Comm.}}} \textbf{\bibinfo{volume}{4}},
  \bibinfo{pages}{1420} (\bibinfo{year}{2013}).

\bibitem{Muirhead16}
\bibinfo{author}{Muirhead, C.~M.}, \bibinfo{author}{Gunupudi, B.} \&
  \bibinfo{author}{Colclough, M.~S.}
\newblock \bibinfo{journal}{\bibinfo{title}{Photon transfer in a system of
  coupled superconducting microwave resonators}}.
\newblock {\emph{\JournalTitle{J. Appl. Phys.}}}
  \textbf{\bibinfo{volume}{120}}, \bibinfo{pages}{084904}
  (\bibinfo{year}{2016}).

\bibitem{Spreeuw90}
\bibinfo{author}{Spreeuw, R. J.~C.}, \bibinfo{author}{van Druten, N.~J.},
  \bibinfo{author}{Beijersbergen, M.~W.}, \bibinfo{author}{Eliel, E.~R.} \&
  \bibinfo{author}{Woerdman, J.~P.}
\newblock \bibinfo{journal}{\bibinfo{title}{Classical realization of a strongly
  driven two-level system}}.
\newblock {\emph{\JournalTitle{Phys. Rev. Lett.}}}
  \textbf{\bibinfo{volume}{65}}, \bibinfo{pages}{2642--2645}
  (\bibinfo{year}{1990}).

\bibitem{Okamoto13}
\bibinfo{author}{Okamoto, H.} \emph{et~al.}
\newblock \bibinfo{journal}{\bibinfo{title}{Coherent phonon manipulation in
  coupled mechanical resonators}}.
\newblock {\emph{\JournalTitle{Nat. Phys.}}} \textbf{\bibinfo{volume}{9}},
  \bibinfo{pages}{480--484} (\bibinfo{year}{2013}).

\bibitem{Deng16}
\bibinfo{author}{Deng, G.-W.} \emph{et~al.}
\newblock \bibinfo{journal}{\bibinfo{title}{Strongly coupled nanotube
  electromechanical resonators}}.
\newblock {\emph{\JournalTitle{Nano Lett.}}} \textbf{\bibinfo{volume}{16}},
  \bibinfo{pages}{5456--5462} (\bibinfo{year}{2016}).

\bibitem{Yamaguchi17}
\bibinfo{author}{Yamaguchi, H.}
\newblock \bibinfo{journal}{\bibinfo{title}{{GaAs}-based micro/nanomechanical
  resonators}}.
\newblock {\emph{\JournalTitle{Semicond. Sci. Technol.}}}
  \textbf{\bibinfo{volume}{32}}, \bibinfo{pages}{103003}
  (\bibinfo{year}{2017}).

\bibitem{Ashhab07}
\bibinfo{author}{Ashhab, S.}, \bibinfo{author}{Johansson, J.~R.},
  \bibinfo{author}{Zagoskin, A.~M.} \& \bibinfo{author}{Nori, F.}
\newblock \bibinfo{journal}{\bibinfo{title}{Two-level systems driven by
  large-amplitude fields}}.
\newblock {\emph{\JournalTitle{Phys. Rev. A}}} \textbf{\bibinfo{volume}{75}},
  \bibinfo{pages}{063414} (\bibinfo{year}{2007}).

\bibitem{Shevchenko12}
\bibinfo{author}{Shevchenko, S.~N.}, \bibinfo{author}{Ashhab, S.} \&
  \bibinfo{author}{Nori, F.}
\newblock \bibinfo{journal}{\bibinfo{title}{Inverse
  {Landau-Zener-St\"uckelberg} problem for qubit-resonator systems}}.
\newblock {\emph{\JournalTitle{Phys. Rev. B}}} \textbf{\bibinfo{volume}{85}},
  \bibinfo{pages}{094502} (\bibinfo{year}{2012}).

\bibitem{Shevchenko06}
\bibinfo{author}{Shevchenko, S.~N.} \& \bibinfo{author}{Omelyanchouk, A.~N.}
\newblock \bibinfo{journal}{\bibinfo{title}{Resonant effects in the strongly
  driven phase-biased {C}ooper-pair box}}.
\newblock {\emph{\JournalTitle{Low Temp. Phys.}}}
  \textbf{\bibinfo{volume}{32}}, \bibinfo{pages}{973--975}
  (\bibinfo{year}{2006}).

\bibitem{Forster14}
\bibinfo{author}{Forster, F.} \emph{et~al.}
\newblock \bibinfo{journal}{\bibinfo{title}{Characterization of qubit dephasing
  by {Landau-Zener-St\"uckelberg-Majorana} interferometry}}.
\newblock {\emph{\JournalTitle{Phys. Rev. Lett.}}}
  \textbf{\bibinfo{volume}{112}}, \bibinfo{pages}{116803}
  (\bibinfo{year}{2014}).

\bibitem{Grossmann91}
\bibinfo{author}{Grossmann, F.}, \bibinfo{author}{Dittrich, T.},
  \bibinfo{author}{Jung, P.} \& \bibinfo{author}{H\"anggi, P.}
\newblock \bibinfo{journal}{\bibinfo{title}{Coherent destruction of
  tunneling}}.
\newblock {\emph{\JournalTitle{Phys. Rev. Lett.}}}
  \textbf{\bibinfo{volume}{67}}, \bibinfo{pages}{516--519}
  (\bibinfo{year}{1991}).

\bibitem{Grifoni98}
\bibinfo{author}{Grifoni, M.} \& \bibinfo{author}{H\"anggi, P.}
\newblock \bibinfo{journal}{\bibinfo{title}{Driven quantum tunneling}}.
\newblock {\emph{\JournalTitle{Phys. Rep.}}} \textbf{\bibinfo{volume}{304}},
  \bibinfo{pages}{229--354} (\bibinfo{year}{1998}).

\bibitem{Miao16}
\bibinfo{author}{Miao, Q.} \& \bibinfo{author}{Zheng, Y.}
\newblock \bibinfo{journal}{\bibinfo{title}{Coherent destruction of tunneling
  in two-level system driven across avoided crossing via photon statistics}}.
\newblock {\emph{\JournalTitle{Sci. Rep.}}} \textbf{\bibinfo{volume}{6}},
  \bibinfo{pages}{28959} (\bibinfo{year}{2016}).

\bibitem{Parafilo18}
\bibinfo{author}{Parafilo, A.~V.} \& \bibinfo{author}{Kiselev, M.~N.}
\newblock \bibinfo{journal}{\bibinfo{title}{Tunable {RKKY} interaction in a
  double quantum dot nanoelectromechanical device}}.
\newblock {\emph{\JournalTitle{Phys. Rev. B}}} \textbf{\bibinfo{volume}{97}},
  \bibinfo{pages}{035418} (\bibinfo{year}{2018}).

\bibitem{Berns06}
\bibinfo{author}{Berns, D.~M.} \emph{et~al.}
\newblock \bibinfo{journal}{\bibinfo{title}{Coherent quasiclassical dynamics of
  a persistent current qubit}}.
\newblock {\emph{\JournalTitle{Phys. Rev. Lett.}}}
  \textbf{\bibinfo{volume}{97}}, \bibinfo{pages}{150502}
  (\bibinfo{year}{2006}).

\end{thebibliography}

\end{document}